\documentclass[12pt,a4paper,epsf]{article}
\usepackage{graphics}
\usepackage{amssymb,amsmath}
\usepackage[dvips]{lscape,graphicx}
\usepackage{cite}
\usepackage{longtable}
\newcommand{\ct}{\cite}
\newcommand{\lb}{\label}

\newcommand{\bc}{\begin{center}}
\newcommand{\ec}{\end{center}}
\newcommand{\bd}{\begin{displaymath}}
\newcommand{\ed}{\end{displaymath}}
\newcommand{\be}{\begin{equation}}
\newcommand{\ee}{\end{equation}}
\newcommand{\ba}{\begin{array}}
\newcommand{\ea}{\end{array}}
\newcommand{\bea}{\begin{eqnarray}}
\newcommand{\eea}{\end{eqnarray}}
\newcommand{\bt}{\begin{tabular}}
\newcommand{\et}{\end{tabular}}
\newcommand{\un}{\underline}

\newcommand{\bp}{\begin{picture}}
\newcommand{\ep}{\end{picture}}
\newcommand{\bfi}{\begin{figure}}
\newcommand{\efi}{\end{figure}}

\def\fun#1#2{\lower3.6pt\vbox{\baselineskip0pt\lineskip.9pt
\ialign{$\mathsurround=0pt#1\hfil##\hfil$\crcr#2\crcr\sim\crcr}}}
\begin{document}
\vspace{1cm}
\title{\LARGE \bf {Graviweak Unification in the Visible and Invisible Universe and Inflation from the Higgs Field False Vacuum}}
\author{\large \bf  C.R.~Das${}^{1}$\footnote
{crdas@cftp.ist.utl.pt, crdas@prl.res.in}\,\, and L.V.~Laperashvili ${}^{2}$\footnote{laper@itep.ru}\\[5mm]
{\large \it ${}^{1}$ Centre for Theoretical Particle Physics (CFTP),
University of Lisbon,}\\
{\large \it Avenida Rovisco Pais, 1 1049-001
Lisbon, Portugal}\\{\large \it and}\\
{\large \it Theoretical Physics Division, Physical Research Laboratory,}\\
{\large \it Navrangpura, Ahmedabad - 380 009, India}\\\\
{\large \it ${}^{2}$ The Institute of Theoretical and
Experimental Physics,}\\
{\large\it National Research Center ``Kurchatov Institute",}\\ {\large
\it Bolshaya Cheremushkinskaya, 25, 117218 Moscow, Russia}}
\date{}
\maketitle
\thispagestyle{empty}

\begin{abstract}

In the present paper we develop the self-consistent
$Spin(4,4)$-invariant model of the unification of gravity with
weak $SU(2)$ interactions in the assumption of the existence of
visible and invisible sectors of the Universe. It was shown that
the consequences of the multiple point principle predicting two
degenerate vacua in the Standard Model (SM) suggest a theory of
Inflation, in which the inflaton field $\sigma$ starts trapped in
a cold coherent state in the ``false vacuum" of the Universe at the
value of the Higgs field's VEV $v\sim 10^{18}$ GeV (in the visible
world). Then the inflations of the two Higgs doublet fields,
visible $\phi$ and mirror $\phi'$, lead to the emergence of the SM
vacua at the Electroweak scales with the Higgs boson VEVs
$v_1\approx 246$ GeV and $v'_1=\zeta v_1$ (with $\zeta \sim 100$)
in the visible and invisible worlds, respectively.

\end{abstract}


\par cosmological constant, dark energy


\thispagestyle{empty}

\clearpage\newpage

\section{Introduction}

In Ref.~\ct {1} a model of unification of gravity with the weak
$SU(2)$ gauge and Higgs fields was constructed, in accordance with
Ref.\ct{2}. Previously gravi-weak and gravi-electro-weak unified
models were suggested in Ref.~\ct{1gw,2gw,3gw}.

In this investigation we imagine that at the early stage of the
evolution of the Universe the GUT-group was broken down to the
direct product of gauge groups of the internal symmetry $U(4)$ and
Spin(4,4)-group of the Graviweak unification.

In the assumption that there exist visible and invisible (hidden)
sectors of the Universe, we presented the hidden world as a Mirror
World (MW) with a broken Mirror Parity (MP). In the present paper
we give arguments that MW is not identical to the visible Ordinary
World (OW). We started with an extended $\mathfrak g =
\mathfrak{spin}(4,4)_L$-invariant Plebanski action in the visible
Universe, and with $\mathfrak g =
\mathfrak{spin}(4,4)_R$-invariant Plebanski action in the MW. Then
we have shown that the Graviweak symmetry breaking leads to the
following sub-algebras: $\tilde {\mathfrak g} = {\mathfrak
sl}(2,C)^{(grav)}_L \oplus {\mathfrak su}(2)_L$ -- in the ordinary
world, and $\tilde {\mathfrak g}' = {{\mathfrak
sl}(2,C)'}^{(grav)}_R \oplus {\mathfrak su}(2)'_R$ -- in the
hidden world. These sub-algebras contain the self-dual left-handed
gravity in the OW, and the anti-self-dual right-handed gravity in
the MW. Finally, at low energies, we obtain a Standard Model (SM)
group of symmetry and the Einstein-Hilbert's gravity. In this
approach we have developed a model of Inflation, in which the
inflaton $\sigma$, being a scalar $SU(2)$-triplet field, decays
into the two Higgs $SU(2)$ doublets of the SM: $\sigma\to
\phi^\dagger \phi$, and then the interaction between the ordinary
and mirror Higgs fields (induced by gravity) leads to the hybrid
model of the Inflation.

In Section 2 we considered the Plebanski's theory of gravity, in
which fundamental fields are 2-forms, containing tetrads, spin
connections and auxiliary fields. Then we have used an extension
of the Plebanski's formalism of the 4-dimensional gravitational
theory, and in Section 3 we constructed the action of the
Graviweak unification model, described by the overall unification
parameter $g_{uni}$. Section 4 is devoted to the Multiple Point
Model (MPM), which  allows the existence of several minima of the
Higgs effective potential with the same energy density (degenerate
vacua). The MPM assumes the existence of the SM itself up to the
scale $\sim 10^{18}$ GeV, and predicts that there exist two
degenerate vacua into the SM: the first one -- at the Electroweak
(EW) scale (with the VEV $v_1\simeq 246$ GeV), and the second one
-- at the Planck scale (with the VEV $v=v_2\sim 10^{18}$ GeV). In
Section 5 we consider the existence in the Nature of the Mirror
World (MW) with a broken Mirror Parity (MP): the Higgs VEVs of the
visible and invisible worlds are not equal,{\footnote {In this
paper the superscript 'prime' denotes the M- or hidden H-world.}
$\langle\phi\rangle=v$, $\langle\phi'\rangle=v'$ and $v\neq v'$.
The parameter characterizing the violation of the MP is $\zeta =
{v'}/{v} \gg 1$. We have used the result $\zeta\simeq 100$. In
Section 6 we suggest a hybrid model of Inflation provided with the
visible Higgs field $\phi$ and mirror Higgs boson $\phi'$, which
interact during Inflation via gravity. This interaction leads to
the emergence of the SM vacua at the EW scales with the Higgs
boson VEVs $v_1\approx 246$ GeV and $v'_1=\zeta v_1$ (with $\zeta
\sim 100$) in the visible and invisible worlds, respectively,
while the original ``false vacuum" existed at the Planck scale and
had VEV $v=v_2\sim 10^{18}$ GeV. Section 7 contains Conclusions.

\section{Plebanski's formulation of General Relativity}

General Theory of Relativity (GTR) was formulated by Einstein as
dynamics of the metrics  $g_{\mu\nu }$. Later, Plebanski \ct{3}
and other authors (see for example \ct{3a,3b}) presented GTR in
the self-dual approach, in which fundamental variables are 1-forms
of connections $A^ {IJ}$ and tetrads $e^I$ :
\be   A^{IJ} =  A_{\mu}^{IJ}dx^{\mu}, \qquad
       e^I = e_{\mu}^Idx^{\mu}.
                          \lb{1} \ee
Also 1-form $A = \frac 12 A^ {IJ}\gamma_ {IJ}$ is used, in which
generators $\gamma_ {IJ}$ are products of generators of the
Clifford algebra $Cl(1,3)$: $\gamma_ {IJ} = \gamma_{I}\gamma_
{J}$. Indices $I, J=0,1,2,3$ belong to the space-time with
Minkowski's metrics $\eta^ {IJ} = {\rm diag} (1,-1,-1,-1)$, which
is considered as a flat space, tangential to the curved space with
the metrics $g_{\mu\nu}$. In this case connection belongs to the
local Lorentz group  $SO(1,3)$, or to the spin group $Spin(1,3)$.
In general case of unifications of gravity with the $SU (N)$ or
$SO (N)$ gauge and Higgs fields (see \ct{2}), the gauge algebra is
$\mathfrak g = \mathfrak {spin}(p, q)$, and we have $I, J = 1,2...
p+q$. In our model of unification of gravity with the weak $SU(2)$
interactions we consider a group of symmetry with the Lie algebra
$\mathfrak {spin}(4,4)$. In this model indices $I,J$ run over all
$8\times 8$ values: $I, J = 1,2..,7,8$.

For the purpose of construction of the action for any unification
theory, the following 2-forms are also considered:
$$ B^{IJ} = e^I\wedge e^J = \frac 12
e_{\mu}^Ie_{\nu}^Jdx^{\mu}\wedge dx^{\nu},\qquad F^{IJ} = \frac 12
F_{\mu\nu}^{IJ}dx^{\mu}\wedge dx^{\nu}, $$ where $F_{\mu\nu}^{IJ}
=\partial_{\mu}A_{\nu}^{IJ} - \partial_{\nu}A_{\mu}^{IJ} +
\left[A_{\mu}, A_{\nu}\right]^{IJ}$, which determines the
Riemann-Cartan curvature: $R_{\kappa \lambda \mu \nu} =
e_{\kappa}^I e_{\lambda}^JF_{\mu\nu}^{IJ}$. Also 2-forms of $B$
and $F$ are considered :
\be B= \frac 12 B^{IJ}\gamma_{IJ}, \qquad  F= \frac 12
F^{IJ}\gamma_{IJ}, \qquad  F = dA + \frac 12 \left[A, A\right].
\lb{9} \ee
The well-known in literature Plebanski's $BF$-theory is submitted
by the following gravitational action with nonzero cosmological
constant $\Lambda$:
\be  I_{(GR)} = \frac{1}{\kappa^2}\int
\epsilon^{IJKL}\left(B^{IJ}\wedge
    F^{KL} + \frac{\Lambda}{4}B^{IJ}\wedge B^{KL}\right),
                                  \lb{11} \ee
where $\kappa^2=8\pi G_N$, $G_N$ is the Newton's gravitational
constant, and $M_{Pl}^{red.} = 1/{\sqrt{8\pi G_N}}$.

Considering the dual tensors: $$F^*_{\mu\nu}\equiv \frac
{1}{2\sqrt{-g}}\epsilon^{\rho\sigma}_{\mu\nu}F_{\rho\sigma}, \quad
A^{\star IJ} = \frac 12 \epsilon^{IJKL}A^{KL},$$ we can determine
self-dual (+) and anti-self-dual (-) components of the tensor $A^
{IJ}$:
\be A^{(\pm)\,IJ}=\left({\cal P}^{\pm}A\right)^{IJ} = \frac 12
\left(A^{IJ} \pm iA^{\star\,IJ}\right).
                                 \lb{14} \ee
Two projectors on the spaces of the so-called self- and
anti-self-dual tensors $${\cal P}^{\pm}= \frac
12\left(\delta^{IJ}_{KL} \pm i\epsilon^{IJ}_{KL}\right)$$ carry
out the following homomorphism:
\be
   \mathfrak{so}(1,3) = \mathfrak{sl}(2,C)_R \oplus
   \mathfrak{sl}(2,C)_L.
                                 \lb{15} \ee
As a result of Eq.~(\ref{15}), non-zero components of connections
are only $A^{(\pm) i } = A^ {(\pm) 0i}$. Instead of
(anti-)self-duality, the terms of left-handed (+) and right-handed
(-) components are used.

Plebanski \ct{3} and other authors \ct{3a,3b} suggested to
consider a gravitational action in the (visible) world as a
left-handed $\mathfrak{sl}(2,C)_L^{(grav)}$- invariant action,
which contains self-dual fields $F=F^{(+)i}$ and
$\Sigma=\Sigma^{(+)i}$ (i=1,2,3):
\be I_{(grav)}(\Sigma,A,\psi) = \frac{1}{\kappa^2} \int
\left[\Sigma^i\wedge F^i +
 \left(\Psi^{-1}\right)_{ij}\Sigma^i\wedge \Sigma^j\right].
                      \lb{18} \ee
Here $\Sigma^i=2B^{0i}$, and $\Psi_{ij}$ are auxiliary fields,
defining a gauge, which provides equivalence of Eq.~(\ref{18}) to
the Einstein-Hilbert gravitational action:
\be I_{(EG)} = \frac{1}{\kappa^2} \int d^4 x (\frac{R}{2} -
\Lambda ),
                           \lb{19} \ee
where $R$ is a scalar curvature, and $\Lambda$ is the Einstein
cosmological constant.

\section{Graviweak unification model}

On a way of unification of the gravitational and weak interactions
we considered an extended $\mathfrak g = \mathfrak
{spin}(4,4)$-invariant Plebanski's action:
\be I(A, B, \Phi) = \frac{1}{g_{uni}} \int_{\mathfrak
M}\left\langle BF +  B\Phi B + \frac 13 B\Phi \Phi \Phi B
\right\rangle, \lb{22} \ee
where $\langle...\rangle$ means a wedge product, $g_{uni}$ is an
unification parameter, and $\Phi_{IJKL}$ are auxiliary fields.

Having considered the equations of motion, obtained by means of
the action (\ref {22}), and having chosen a possible class of
solutions, we can present the following action for the Graviweak
unification (see details in Refs.~\ct{1,2}):
\be   I(A, \Phi) = \frac{1}{8g_{uni}} \int_{\mathfrak M} \langle
\Phi FF \rangle, \lb{38} \ee
where
\be  \langle \Phi F F \rangle =
\frac{d^4x}{32}\epsilon^{\mu\nu\rho\sigma}{{\Phi_{\mu\nu}}^{\varphi\chi
IJ}}_{KL}F_{\varphi\chi IJ} {F_{\rho\sigma}}^{KL},  \lb{39} \ee
and
\be {{\Phi_{\mu\nu}}^{\rho\sigma ab}}_{cd} = (e_{\mu}^f)
(e_{\nu}^g){\epsilon_{fg}}^{kl}(e^{\rho}_k)
(e^{\sigma}_l)\delta_{cd}^{ab}. \lb{41} \ee
A spontaneous symmetry breaking of our new action that produces
the dynamics of gravity, weak $SU(2)$ gauge and Higgs fields,
leads to the conservation of the following sub-algebra:
$$  \tilde {\mathfrak g} = {{\mathfrak sl}(2,C)}^{(grav)}_L
\oplus {\mathfrak su}(2)_L.$$ Considering indices $a, b \in
\{0,1,2,3\}$ as corresponding to $I,J=1,2,3,4$, and indices $m, n$
as corresponding to indices $I, J=5,6,7,8$, we can present a
spontaneous violation of the Graviweak unification symmetry in
terms of the 2-forms: $$A = \frac 12 \omega + \frac 14 E + A_W,$$
where $\omega = \omega^{ab}\gamma_{ab}$ is a gravitational
spin-connection, which corresponds to the sub-algebra $\mathfrak
{sl}(2,C)_L^{(grav)}$. The connection $E = E^{am}\gamma_{am}$
corresponds to the non-diagonal components of the matrix $A^{IJ}$,
described by the following way (see \ct{2}): $E = e\varphi =
e^a_{\mu}\gamma_a\varphi^m\gamma_m dx^{\mu}$. The connection $A_W
= \frac 12 A^{mn}\gamma_{mn}$ gives: $A_W = \frac 12 A_W^i
\tau_i$, which corresponds to the sub-algebra $\mathfrak
{su}(2)_L$ of the weak interaction. Here $\tau_i$ are the Pauli
matrices with $i=1,2,3$.

Assuming that we have only scalar field
$\varphi^m=(\varphi,\varphi^i)$, we can consider a symmetry
breakdown of the Gravi-Weak Unification, leading to the following
OW-action \ct{1}:
$$I_{(OW)}\left(e,\varphi,A,A_W\right)= \frac{3}{8g_{uni}}
\int_{\bf M} d^4x|e|\left(\frac 1{16}{|\varphi|}^2 R -
\frac{3}{32}{|\varphi|}^4 \right.$$
 \be + \frac 1{16}{R_{ab}}^{cd}
{R^{ab}}_{cd} - \left.\frac 12 {\cal D}_a{\varphi^{\dagger}} {\cal
D}^a\varphi - \frac 14 {F_W^i}_{ab}{F_W^i}^{ab} \right). \lb{27u}
\ee
In Eq.~(\ref{27u}) we have the Riemann scalar curvature $R$;
$|\varphi |^2 = {\varphi }^{\dag}\varphi$ is a squared scalar
field, which from the beginning is not the Higgs field of the
Standard Model; ${\cal D }\varphi = d\varphi + [A_W, \varphi]$ is
a covariant derivative of the scalar field, and $F_W = dA_W +
[A_W, A_W]$ is a curvature of the gauge field $A_W$. The third
member of the action (\ref{27u}) is a topological term in the
Gauss-Bone theory of gravity (see for example \ct{4,4a}).

Lagrangian in the action (\ref{27u}) leads to the nonzero vacuum
expectation value (VEV) of the scalar field:
$v=\langle\varphi\rangle =\varphi_0$, which corresponds to a local
minimum of the effective potential $V_{eff} (\varphi)$  at $v^2 =
R_0/3$, where $R_0 > 0$ is a constant de Sitter space-time
background curvature \ct{2}.

According to (\ref{27u}), the Newton gravitational constant $G_N$
is defined by the expression:
\be 8\pi G_N = ({M^{(red.)}_{Pl}})^{-2} = \frac{64g_{uni}}{3v^2},
\lb{28u} \ee
a bare cosmological constant is equal to
\be \Lambda_0 = \frac 34 v^2, \lb{29u} \ee
and
\be g_W^2 = 8g_{uni}/3. \lb{30u} \ee
The coupling constant $g_W$ is a bare coupling constant of the
weak interaction, which also coincides with a value of the
constant $g_2=g_W$ at the Planck scale. Considering the running
$\alpha_2^ {-1}(\mu)$, where $\alpha_2=g_2^2/4\pi$, we can carry
out an extrapolation of this rate to the Planck scale, what leads
to the following estimation \ct{5,5a}:
\be \alpha_2 (M_{Pl}) \sim 1/50,  \lb{31u} \ee
and then the overall GWU parameter is: $g_{uni}\sim 0.1.$

\section{Multiple Point Model}

The radiative corrections to the effective Higgs potential,
considered in Refs.~\ct{6,7}, bring to the emergence of the second
minimum of the effective Higgs potential at the Planck scale. It
was shown that in the 2-loop approximation of the effective Higgs
potential, experimental values of all running coupling constants
in the SM predict an existence of the second minimum of this
potential located near the Planck scale, at the value $v_2
=\varphi_{min2}\sim M_{Pl}$.

In general, a quantum field theory allows an existence of several
minima of the effective potential, which is a function of a scalar
field. If all vacua, corresponding to these minima, are
degenerate, having zero cosmological constants, then we can speak
about the existence of a multiple critical point (MCP) at the
phase diagram of theory considered for the investigation (see
Refs.~\ct{8,8a}).  In Ref.~\ct{8} Bennett and Nielsen suggested
the Multiple Point Model (MPM) of the Universe, which contains
simply the SM itself up to the scale $\sim 10^{18}$ GeV. In
Ref.~\ct{9} the MPM was applied (by the consideration of the two
degenerate vacua in the SM) for the prediction of the top-quark
and Higgs boson masses, which gave:
\be M_t = 173 \pm 5 \,\, {\rm GeV }, \qquad M_H = 135 \pm 9 \,\,
{\rm GeV }. \lb{29} \ee
Later, the prediction for the mass of the Higgs boson was improved
by the calculation of the two-loop radiative corrections to the
effective Higgs potential \ct{6,7}. The predictions: 125 GeV
$\lesssim M_H \lesssim$ 143 GeV in Ref.~\ct {6}, and 129 $\pm$ 2
GeV in Ref.~\ct{7} -- provided the possibility of the theoretical
explanation of the value $M_H\approx$ 126 GeV observed at the LHC.
The authors of Ref.~\ct{7a} have shown that the most interesting
aspect of the measured value of $M_H$ is its near-criticality.
They have thoroughly studied the condition of near-criticality in
terms of the SM parameters at the high (Planck) scale. They
extrapolated the SM parameters up to large energies with full
3-loop NNLO RGE precision. All these results mean that the
radiative corrections to the Higgs effective potential lead to the
value of the Higgs mass existing in the Nature.

Having substituted in Eq.(\ref{28u}) the values of $g_ {uni}
\simeq 0.1$ and $G_N=1/8\pi {(M_{Pl}^{red.})}^2$, where $M_{Pl}^{
red.}\approx 2.43\cdot 10^{18}$ GeV, it is easy to obtain the
VEV's value $v$, which in this case is located near the Planck
scale:
\be v=v_2\approx 3.5\cdot 10^{18} {\rm{GeV}}. \lb{32u} \ee
Such a result takes place, if the Universe at the early stage
stayed in the "false vacuum", in which the VEV of the Higgs field
is huge: $v=v_2\sim 10^{18} GeV$. The exit from this state could
be carried out only by means of the existence of the second scalar
field. In the present paper we assume that the second scalar
field, participating into the Inflation, is the mirror Higgs
field, which arises from the interaction between the Higgs fields
of the visible and invisible sectors of the Universe.

\section{Mirror world with broken mirror parity}

As it was noted at the beginning of this paper, we assumed the
parallel existence in the Nature of the visible (OW) and invisible
(MW) (mirror) worlds.

Such a hypothesis was suggested in Refs.~\ct{LY,KOP}.

The Mirror World (MW) is a mirror copy of the Ordinary World (OW)
and contains the same particles and types of interactions as our
visible world, but with the opposite chirality. Lee and Yang
\ct{LY} were first to suggest such a duplication of the worlds,
which restores the left-right symmetry of the Nature. The term
``Mirror Matter'' was introduced by Kobzarev, Okun and Pomeranchuk
\ct{KOP}, who first suggested to consider MW as a hidden
(invisible) sector of the Universe, which interacts with the
ordinary (visible) world only via gravity, or another (presumably
scalar) very weak interaction.

In the present paper we consider the hidden sector of the Universe
as a Mirror World (MW) with broken Mirror Parity (MP)
\ct{1mw,2mw,3mw,4mw,5mw}. If the ordinary and mirror worlds are
identical, then O- and M-particles should have the same
cosmological densities. But this is immediately in conflict with
recent astrophysical measurements \ct{1D,2D,3D}. Astrophysical and
cosmological observations have revealed the existence of the Dark
Matter (DM), which constitutes about 25\% of the total energy
density of the Universe. This is five times larger than all the
visible matter, $\Omega_{DM}: \Omega_{M} \simeq 5 : 1$. Mirror
particles have been suggested as candidates for the inferred dark
matter in the Universe \ct{4mw,6mw,7mw} (see also \ct{Sil}).
Therefore, the mirror parity (MP) is not conserved, and the OW and
MW are not identical.

The group of symmetry $G_ {SM}$ of the Standard Model was enlarged
to $G_ {SM}\times G'_ {SM'}$, where $G_{SM}$ stands for the
observable SM, while $ G'_ {SM'}$ is its mirror gauge counterpart.
Here O(M)- particles are singlets of the group $G_ {SM}$ ($G _
{SM'}$).

It was assumed that the VEVs of the Higgs doublets $\phi$ and
$\phi'$ are not equal:
$$\langle\phi\rangle=v, \quad \langle\phi' \rangle=v', \quad {\rm
{and} }\quad v\neq v'. $$ The parameter characterizing the
violation of the MP is $\zeta = {v'}/{v} \gg 1$. Astrophysical
estimates give: $\zeta > 30, \quad \zeta \sim 100$ (see references
in \ct{10mw,11mw}).

The action $I_{(MW)}$ in the mirror world is represented by the
same integral (\ref{27u}), in which we have to make the
replacement of all OW-fields by their mirror counterparts: $e,
\phi, A, A_W, R \to e', \phi', A', A' _W, R'$. Then: \be G' _N
=\zeta G_N,\quad \Lambda'_0 = \zeta^2\Lambda_0, \quad {M'} _ {Pl}^
{red.} = \zeta M_{Pl}^{red.}.  \lb{29}\ee However, $g' _W = g_W$:
it is supposed that at the early stage of evolution of the
Universe, when the GUT takes place, mirror parity is unbroken,
what gives $g'_{uni} = g_{uni}$.

\section{Inflation model}

It is well-known that the hidden (invisible) sector of the
Universe interacts with the ordinary (visible) world only via
gravity, or another very weak interaction (see for example
\ct{KOP,KST,7mw}). In particular, the authors of Ref.~\ct{12mw}
assumed, that along with gravitational interaction there also
exists the interaction between the initial Higgs fields of both
OW- and MW-worlds:
\be V_{int} = \alpha_h( \varphi^{\dagger} \varphi)( {\varphi'}
^{\dagger}{\varphi'}), \lb{30} \ee
which begin to interact during the Inflation via gravitational
interactions. The existence of the second Higgs field $\varphi'$
could be the cause of the hybrid inflation (see the model of
Hybrid inflation by A.~Linde \ct{Lin}), bringing the Universe out
of the ``false vacuum" with the VEV $v_2\sim 10^{18}$ GeV. This
circumstance provided the subsequent transition to the vacuum with
the Higgs VEV $v_1$ existing at the Electroweak (EW) scale. Here
$v_1\approx 246$ GeV is a vacuum, in which we live at the present
time.

In Section 3 we obtained the GWU action given by Eq.~(\ref{27u}).
The gravitational part of the action is:
\be  I_{(OW)}\left(e,\varphi,A,A_W\right)
 = \frac{3}{64g_{uni}}
\int_{\bf M} d^4x|e|\left(\frac 12{|\varphi|}^2 R -
\frac{3}{4}{|\varphi|}^4
 +...\right).  \lb{1z} \ee
Considering the background value $R\simeq R_0$, we can find a
minimum of the potential:
\be    V_{eff}(\varphi) \sim - \frac 12{|\varphi|}^2 R_0 +
\frac{3}{4}{|\varphi|}^4
 \lb{2z} \ee
at $\varphi_0=<\varphi>=v$. Here $v^2=R_0/3$. Then according to
(\ref{28u}), we obtain:
\be   I_{(OW)}\left(e,\varphi,A,A_W\right)= \int_{\bf M}
d^4x\sqrt{-g}\left(\frac{M^{red}_{Pl}}{v}\right)^2\left(\frac
12{|\varphi|}^2R - \frac{3}{4}{|\varphi|}^4
 +...\right).  \lb{3z} \ee
In the action (\ref{3z}) the Lagrangian includes the non-minimal
coupling with gravity \ct{1Bez,2Bez,3Bez}.

We see that the field $\varphi$ is not stuck at $\varphi_0$
anymore, but it can be represented as
\be \varphi = \varphi_0 - \sigma = v - \sigma ,  \lb{4z} \ee
where the scalar field $\sigma$ is an {\un {inflaton}}. Here we
see that in the minimum, when $\varphi=v$, the inflaton field is
zero ($\sigma=0$) and then it increases with falling of the field
$\varphi$.

Considering the expansion of the Lagrangian around the background
value $R\simeq R_0$ in powers of the small value $\sigma/v$, and
leaving only the first-power terms, we can present the
gravitational part of the action as:
\be I_{(grav\,\, OW)} = \int_{M}d^4x
\sqrt{-g}\left(M_{Pl}^{red}\right)^2 \left( \frac 12 R_0 |1 -
\sigma/v|^2 - \Lambda_0|1 - \sigma/v|^4 + ...\right). \lb{5z} \ee
Here $\Lambda_0 = \frac 34 v^2 = R_0/4$. Using the last relations,
we obtain:
\be I_{(grav\,\, OW)} = \int_{M}d^4x
\sqrt{-g}\left(M_{Pl}^{red}\right)^2 \left( \Lambda_0 -
\frac{m^2}{2}|\sigma|^2 + ...\right), \lb{6z} \ee
where $m^2 = 6$ and $m$ is the bare mass of the inflaton in units
$M_{Pl}^{red.}=1$.

In the Einstein-Hilbert action the vacuum energy is:
\be \rho_{vac} = \left(M_{Pl}^{red}\right)^2 \Lambda. \lb{7z} \ee
In our case (\ref{6z}) the vacuum energy density is negative:
 \be \rho_0 =
- \left(M_{Pl}^{red}\right)^2 \Lambda_0. \lb{8z} \ee
However, assuming the existence of the discrete space-time of the
Universe at the Planck scale and using the prediction of the
non-commutativity suggested  by B.G. Sidharth \ct{1S,2S}, we
obtain that the gravitational part of the GWU action has the
vacuum energy density equal to zero or almost zero.

Indeed, the total cosmological constant and the total vacuum
density of the Universe contain also the vacuum fluctuations of
fermions and other SM boson fields:
\be \Lambda \equiv \Lambda_{eff} = \Lambda^{ZMD} - \Lambda_0 -
\Lambda^{(NC)}_s + \Lambda^{(NC)}_f, \lb{9z} \ee
where $\Lambda^{ZMD}$ is zero modes degrees of freedom of all
fields existing in the Universe, and $\Lambda^{(NC)}_{s,f}$ are
boson and fermion contributions of the non-commutative geometry of
the discrete spacetime at the Planck scale. If according to the
theory by B.G. Sidharth \ct{1S,2S}, we have:
\be \rho_{vac}^{(0)} = \left(M_{Pl}^{red}\right)^2 \Lambda^{(0)} =
 \left(M_{Pl}^{red}\right)^2 (\Lambda^{ZMD} -
\Lambda_0 - \Lambda_s^{(NC)}) \approx 0, \lb{10z} \ee
then Eq.~(\ref{6z}) contains the cosmological constant
$\Lambda^{(0)} \approx 0$. In Eqs.~(\ref{9z}) and (\ref{10z}) the
bosonic (scalar) contribution of the non-commutativity is:
\be \rho_{(scalar)}^{(NC)} = m_s^4 \quad ({\rm{in \, units}}:\,
\hbar = c = 1), \lb{11z} \ee
which is given by the mass $m_s$ of the primordial scalar field
$\varphi$. Then the discrete spacetime at the very small distances
is a lattice (or has a lattice-like structure) with a parameter $a
= \lambda_s = 1/m_s.$ This is a scalar length:
$$
a = \lambda_s \sim 10^{-19}\,\,{\rm GeV}^{-1},$$ which coincides
with the Planck length: $\lambda_{Pl} = 1/M_{Pl} \approx
10^{-19}\,\,{\rm GeV}^{-1}.$
The assumption:
\be \Lambda^{(0)} = \Lambda^{ZMD} - \Lambda_0 - \Lambda_s^{(NC)}
\approx 0 \lb{12z} \ee
means that the Gravi-Weak Unification model contains the
cosmological constant equal to zero or almost zero.

B.G.~Sidharth gave in Ref.~\ct{11S} the estimation:
\be \rho_{DE} = \left(M_{Pl}^{red}\right)^2 \Lambda_f^{(NC)},
\lb{13z} \ee
considering the non-commutative contribution of light primordial
neutrinos as a dominant contribution to $\rho_{DE}$, which
coincides with astrophysical measurements \ct{1D,2D,3D}:
\be \rho_{DE} \approx (2.3\times 10^{-3}\,\,\rm{eV})^4. \lb{14z}
\ee
Returning to the Inflation model, we rewrite the action (\ref{6z})
as:
\be I_{(grav\,\, OW)} = - \int_{\large M} d^4x \sqrt{-g}
\left(M_{Pl}^{red}\right)^2 \left(\Lambda + \frac{m^2}{2}
|\sigma|^2 + ...\right), \lb{15z} \ee
where the positive cosmological constant $\Lambda$ is not zero,
but is very small.

Taking into account the interaction of the ordinary and mirror
scalar bosons $\varphi$ and $\varphi'$, given by equation
analogous to Eq.~(\ref{30}) \ct{12mw}, we obtain:
$$I_{(grav)} = \int_{\large M} d^4x \sqrt{-g}\left[
\left(\frac{M_{Pl}^{red}}{v}\right)^2 \left (\frac 12{|\varphi|}^2
R - \frac{3}{4}{|\varphi|}^4
-\alpha_h{|\varphi|}^2{|{\varphi}'|}^2
 +...\right)+ ... \right ]$$
 \be +  \int_{\large M} d^4x \sqrt{-g}\left[\left(\frac{{ M'}_{Pl}^{red}}
 { v'}\right)^2 \left (\frac 12{|\varphi'|}^2
{R'} - \frac{3}{4}{|\varphi'|}^4
-\alpha_h{|\varphi|}^2{|\varphi'|}^2
 +...\right) + ... \right].   \lb{18z} \ee
Considering the Planck scale Higgs potential, corresponding to the
action (\ref{18z}), we have:
$$ V(\varphi, \varphi') \simeq
\left(\frac{M_{Pl}^{red}}{v}\right)^2 \left ( - \frac
12{|\varphi|}^2 R_0  + \frac{3}{4}{|\varphi|}^4
 + \alpha_h{|\varphi|}^2{|\varphi'|}^2 \right)$$ \be +
\left(\frac{{M'}_{Pl}^{red}}
 { v'}\right)^2 \left( - \frac 12{|\varphi'|}^2
R'_0 + \frac{3}{4}{|\varphi'|}^4 +
{\alpha}_h{|\varphi|}^2{|\varphi'|}^2 \right). \lb{19z} \ee
According to (\ref{28u}), we have:
\be  \left(\frac{M_{Pl}^{red}}{v}\right)^2 =
\left(\frac{{M'}_{Pl}^{red}}{v'}\right)^2, \lb{20z} \ee
and the local minima at $\varphi_0=v$ and ${\varphi'}_0 =  v'$ are
given by the following conditions:
\be \frac{\partial V(\varphi,
\varphi')}{\partial{|\varphi|^2}}\big|_{|\varphi|=v} =
\left(\frac{M_{Pl}^{red}}{v}\right)^2 \left( - \frac 12 R_0  +
\frac{3}{2}v^2
 + 2\alpha_h{|\varphi'|}^2 \right) = 0,
    \lb{22z} \ee
 \be \frac{\partial V(\varphi,
\varphi')}{\partial{|\varphi'|^2}}\big|_ {|\varphi'|=v'} =
\left(\frac{M_{Pl}^{red}}{v}\right)^2 \left( - \frac 12  R'_0 +
\frac{3}{2}{v'}^2 + 2\alpha_h{|\varphi|}^2 \right) = 0,
    \lb{23z} \ee
which give the following solutions:
 \be  v^2 \simeq \frac{R_0}{3} - \frac 43
 \alpha_h{|\varphi'|}^2,
                            \lb{24z} \ee
 \be {v'}^2 \simeq \frac{R'_0}{3} - \frac
 43\alpha_h{|\varphi|}^2,
                        \lb{25z} \ee
and
 \be  V(\varphi = v, \varphi' = v') = - \frac 14
\big[ (M_{Pl}^{red})^2R_0 + ({M'}_{Pl}^{red})^2
 R'_0 \big] = - (M_{Pl}^{red})^2\Lambda_0 -
 ({M'}_{Pl}^{red})^2{\Lambda'}_0.
                            \lb{26z} \ee
Finally, according to (\ref{28u}), we obtain:
 \be  V(\varphi = v, \varphi' = v') =
= - (1 + \zeta^4)(M_{Pl}^{red})^2\Lambda_0,
                            \lb{27z} \ee
what gives the negative vacuum energy density.

However, the cosmological constant is not given by
Eq.~(\ref{27z}). According to the ideas of non-commutativity given
by B.G.~Sidharth in Refs.~\ct{1S,2S,11S}, it must be replaced by
the cosmological constant $\Lambda$, which is related with the
potential $V(\varphi = v,\,\, \varphi' = v')$ and with the Dark
Energy density (\ref{14z}) by the following way:
 \be  V(\varphi = v,\,\, \varphi' = v') =
(M_{Pl}^{red})^2\Lambda = \rho_{DE},
                    \lb{28z} \ee
where
$$ \Lambda =  \Lambda_{eff}  +   \Lambda'_{eff}, $$
$\Lambda_{eff}$ is given by Eq.~(\ref{9z}), and $\Lambda'_{eff}$
is its mirror counterpart.

Then using the notation:
 \be \varphi = v - \sigma \quad {\rm{and}}\quad
\varphi' = v' - {\sigma}', \lb{29z} \ee
and neglecting the terms containing $\alpha_h$ as very small, it
is not difficult to see that the potential near the Planck scale
is:
 \be  V(\varphi, \,\,\varphi') =
(M_{Pl}^{red})^2(\Lambda + \frac{m^2}2 |\sigma|^2 + \frac{
{m'}^2}{2}|{\sigma}'|^2 + ...),  \lb{30z} \ee
where $m^2 \simeq 6$ and ${m'}^2 \simeq 6\zeta^2$ (compare with
(\ref{14z})).

The local minimum of the potential (\ref{30z}) at $\varphi_0=v$,
when $\sigma=0$, and ${\varphi'}_0 \neq v'$ ($\sigma' \neq 0$)
gives:
 \be  V( v,\,\, \varphi')= {(M_{Pl}^{red})}^2(\Lambda + \frac{
{m'}^2}2|\sigma'|^2 + ...).
                           \lb{31z} \ee
The last equation (\ref{31z}) shows that the potential $V(v)$
grows with growth of $\sigma'$, i.e. with falling of the field
$\varphi'$. It means that a barrier of potential grows and at some
value $\sigma' = \sigma'|_{in}$ potential begins its inflationary
falling. Here it is necessary to comment that the position of the
minimum also is displaced towards smaller $\varphi$ (bigger
$\sigma$), according to the formula (\ref{24z}).

Our next step is an assumption that during the inflation $\sigma$
decays into the two Higgs doublets of the SM:
\be
      \sigma \to \phi^{\dagger} + \phi.   \lb{32z} \ee
As a result, we have:
 \be \sigma = a_d|\phi|^2, \lb{33z} \ee
where $\phi$ is the Higgs doublet field of the Standard Model. The
Higgs field $\phi$ also interacts directly with field $\phi'$,
according to the interaction (\ref{30}) given by Ref.~\ct{12mw}.
It has a time evolution and modifies the shape of the barrier so
that at some value $\phi'_E$ can roll down the field $\varphi$.
This possibility, which we consider in our paper, is given by the
so-called Hybrid Inflation scenarios \ct{Lin}. Here we assume that
the field $\phi$ begins the inflation at the value
$\phi|_{in}\simeq H_0$.

Using the relations given by GWU, we obtain near the local ``false
vacuum'' the following gravitational potential:
\be V(\phi,\,\,{\phi'}) \simeq \Lambda + \frac{\lambda}4 |\phi|^4
+ \frac{\lambda'}4|{\phi'}|^4 + \frac{a_h}4|\phi|^2|{\phi'}|^2,
\lb{34z} \ee
where $\lambda=12a_d$  and ${\lambda}' = 12{(a'_d)}^2$ are
self-couplings of the Higgs doublet fields $\phi$ and ${\phi}'$,
respectively.

Returning to the problem of the Inflation, we see that the action
of the GWU theory has to be written near the Planck scale as:
\be I_{grav} \simeq - \int_M d^4x \sqrt{-g}
\left(M_{Pl}^{red}\right)^2 \left(\Lambda + \frac{\lambda}4
|\phi|^4 + \frac{\lambda'}4|{\phi'}|^4 +
\frac{a_h}4|\phi|^2|{\phi'}|^2
 + ...\right), \lb{35z} \ee
where the cosmological constant $\Lambda$ is almost zero (has an
extremely tiny value).

The next step is to see the evolution of the Inflation in our
model, based on the GWU with two Higgs fields $\phi$ and mirror
$\phi'$.

In the present investigation we considered only the result of such
an Inflation, which corresponds to the assumption of the MPP, that
cosmological constant is zero (or almost zero) at both vacua:  at
the "first vacuum" with VEV $v_1 = 246$ GeV and at the "second
vacuum" with VEV $v=v_2\sim 10^{18}$ GeV. If so, we have the
following conditions of the MPP (see section 4):\\
\be V_{eff} \left(\phi_{min 1}\right) = V_{eff} \left(\phi_{min
2}\right) = 0, \lb{20zz} \ee
\be  \frac{\partial V_{eff}}{\partial |\phi|^2} \big|_ {\phi =
\phi_{min 1}} = \frac{\partial V_{eff}}{\partial |\phi|^2} \big|_
{\phi = \phi_{min 2}} = 0. \lb{36z} \ee
Considering the total Universe as two worlds, ordinary OW and
mirror MW, we present the following expression for the low energy
total effective Higgs potential (which is far from the Planck
scale):
 \be V_{eff} = - \frac{\mu^2}2 |\phi|^2 + \frac 14\lambda(\phi)|\phi|^4
 - \frac{{\mu'}^2}2 |{\phi'}|^2 + \frac 14\lambda'(\phi')|\phi'|^4
 + \frac 14\alpha_h (\phi, \phi') |\phi|^2 |{\phi'}|^2,
\lb{37z} \ee
where $\alpha (\phi, \phi')$ is a coupling constant of the
interaction of the ordinary Higgs field $\phi$ with mirror Higgs
field $\phi'$.

According to the MPP, at the critical point of the phase diagram
of our theory, corresponding to the "second vacuum", we have:
\be \mu = \mu' = 0, \quad \lambda (\phi_0) \simeq 0, \quad
{\lambda'}({\phi'}_0) \simeq 0, \lb{38z} \ee
and then \be \alpha_h (\phi_0, {\phi'}_0) \simeq 0,\quad
{\rm{if}}\quad V_{eff}^{crit} (v_2) \simeq 0.  \lb{39z} \ee
At the critical point, corresponding to the first EW vacuum
$v_1=246$ GeV, we also have $V_{eff}^{crit} (v_1) \simeq 0$,
according to the MPP prediction of the existence of the almost
degenerate vacua in the Universe.

Then we can present the full scalar Higgs potential by the
following expression:
\be V_{eff} (\phi,{\phi}') = \frac 14\big(\lambda {(|\phi|^2 -
v^2_1)}^2 + {\lambda'} {(|\phi'|^2 - {v'}_1^2)}^2 + \alpha_h
(\phi, \phi') (|\phi'|^2 - {v'}_1^2) |\phi|^2\big), \lb{25zz} \ee
where we have shifted the interaction term: \be V_{int} =\frac 14
\alpha_h(\phi,{\phi'}) |\phi|^2 |{\phi'}|^2     \lb{26zz} \ee
in such a way that the interaction term vanishes, when ${\phi'} =
{\phi'}_0 = {v'}_1$, recovering the usual Standard Model.

At the end of the Inflation we have: $\phi' = {\phi'}_E$, and the
first vacuum value of $V_{eff}$ is given by:
\be V_{eff} (v_1, {\phi'}_E) = \frac 14\big(
{\lambda'}(|{\phi'}_E|^2 - {v'}_1^2)^2
 + \alpha_h (v_1,{\phi'}_E) (|{\phi'}_E|^2 -
{v'}_1^2) v_1^2\big) = 0, \lb{27zz} \ee
and
\be \frac{\partial V_{eff}}{\partial |\phi|^2}
\left|\begin{array}{l}
\phi = v_1\\
{\phi'} = {\phi'}_E\end{array}\right| = 0.
                       \lb{28zz} \ee
This means that the end of the Inflation occurs at the value:
\be {\phi'}_E = {v'}_1 = \zeta v_1, \lb{29zz} \ee
which coincides with the VEV \,$<{\phi'}>$\,  of the field
${\phi'}$ at the first vacuum in the mirror world MW. Thus,
\be V_{eff} (\phi,{\phi'}_E) = \frac14 \lambda {(|\phi|^2 - v^2_1)}^2,
\lb{30zz} \ee
which means the Standard Model with the first vacuum, having the
VEV $v_1\approx 246 $ GeV.

\section{Conclusions}

In the present paper we constructed a model of unification of
gravity with the weak $SU(2)$ gauge and Higgs fields. Imagining
that at the early stage of the evolution  the Universe was
described by a GUT-group, we assumed that this Grand Unification
group of symmetry was quickly broken down to the direct product of
the gauge groups of internal symmetry and Spin(4,4)-group of the
Graviweak unification.

Also we assumed the existence of visible and invisible (hidden)
sectors of the Universe. We have given arguments that modern
astrophysical and cosmological measurements lead to a model of the
Mirror World with a broken Mirror Parity (MP), in which the Higgs
VEVs of the visible and invisible worlds are not equal:
$\langle\phi\rangle=v, \quad \langle\phi'\rangle=v' \quad {\rm
{and}}\quad v\neq v'$. We estimated a parameter characterizing the
violation of the MP: $\zeta = v'/v \gg 1$. We have used the
result: $\zeta \sim 100$ obtained by Z.~Berezhiani and his
collaborators. In this model, we showed that the action for
gravitational and $SU(2)$ Yang--Mills and Higgs fields,
constructed in the ordinary world (OW), has a modified duplication
for the hidden (mirror) sector of the Universe (MW).

Considering the Graviweak symmetry breaking, we have obtained the
following sub-algebras: $\tilde {\mathfrak g} = {\mathfrak
su}(2)^{(grav)}_L \oplus {\mathfrak su}(2)_L$ -- in the ordinary
world, and $\tilde {\mathfrak g}' = {{\mathfrak
su}(2)'}^{(grav)}_R \oplus {\mathfrak su}(2)'_R$ -- in the hidden
world. These sub-algebras contain the self-dual left-handed
gravity in the OW, and the anti-self-dual right-handed gravity in
the MW. We assumed, that finally at low energies, we have a
Standard Model and the Einstein-Hilbert's gravity.

We reviewed the Multiple Point Model (MPM) by D.L. Bennett and
H.B.Nielsen, who assumed the existence of several minima of the
Higgs effective potential with the same energy density (degenerate
vacua of the SM). In the assumption of zero cosmological
constants, MPM postulates that all the vacua, which might exist in
the Nature (as minima of the effective potential), should have
zero, or approximately zero, cosmological constant.  The
prediction that there exist two vacua into the SM: the first one
-- at the Electroweak scale ($v_1\simeq 246$ GeV), and the second
one -- at the Planck scale ($v_2\sim 10^{18}$ GeV), was confirmed
by calculations in the 2-loop approximation of the Higgs effective
potential. The prediction of the top-quark and Higgs masses was
given in the assumption that there exist two vacua into the SM.

In the above-mentioned theory we have developed a model of
Inflation. According to this model, a singlet field $\sigma$,
being an inflaton, starts trapped from the ``false vacuum" of the
Universe at the value of the Higgs field's VEV $v =v_2 \sim
10^{18}$ GeV. Then during the Inflation $\sigma$ decays into the
two Higgs doublets of the SM: $\sigma\to \phi^\dagger \phi$. The
interaction between the ordinary and mirror Higgs fields $\phi$
and $\phi'$, induced by gravity, generates a hybrid model of the
Inflation in the Universe. Such an interaction leads to the
emergence of the SM vacua at the Electroweak scales: with the
Higgs boson VEVs $v_1\approx 246$ GeV -- in the OW, and
$v'_1=\zeta v_1$ -- in the MW.

\section{Acknowledgments}

 L.V.~Laperashvili greatly thanks the Niels Bohr Institute
(Copenhagen, Denmark) and Prof. H.B.~Nielsen for hospitality,
collaboration and financial support. L.V.L. also deeply thanks the
Department of Physics and University of Helsinki for hospitality
and financial support, and Prof. M.~Chaichian and Dr A.~Tureanu
for fruitful discussions and advises.

C.R.~Das acknowledges a scholarship
from the Funda\c{c}\~{a}o para a Ci\^{e}ncia e a Tecnologia (FCT,
Portugal) (ref. SFRH/BPD/41091/2007), and greatly thanks the Department of Physics,
Jyv\"askyl\"{a} University, in particular Prof. Jukka Maalampi (HOD) for
hospitality and financial support.
This work was partially
supported by FCT through the projects CERN/FP/123580/2011,
PTDC/FIS-NUC/0548/2012 and CFTP-FCT Unit 777 (PEst-OE/FIS/UI0777/2013)
which are partially funded through POCTI (FEDER). C.R.~Das also sincerely thanks
Physical Research Laboratory and Prof. Utpal Sarkar (Dean) for Visiting Scientist position.

\end{document}